\documentclass{ws-ijmpa}

\begin{document}

\markboth{R.~Czy{\.z}ykiewicz et al.}
{Production of $\eta$ mesons in proton-proton collisions close to threshold}

%
\catchline{}{}{}{}{}
%

\title{PRODUCTION OF $\eta$ MESONS
IN PROTON-PROTON COLLISIONS CLOSE TO THRESHOLD
}

\author{\footnotesize R.~CZY{\.Z}YKIEWICZ$^{\alpha,\beta}$, H-H.~ADAM$^{\gamma}$, 
A.~BUDZANOWSKI$^{\delta}$, D.~GRZONKA$^{\beta}$, M.~JANUSZ$^{\alpha}$, L.~JARCZYK$^{\alpha}$,
B.~KAMYS$^{\alpha}$, A.~KHOUKAZ$^{\gamma}$, K.~KILIAN$^{\beta}$, P.~KLAJA$^{\alpha}$, P.~KOWINA$^{\epsilon}$, P.~MOSKAL$^{\alpha,\beta}$, 
W.~OELERT$^{\beta}$, C.~PISKOR-IGNATOWICZ$^{\alpha}$, J.~PRZERWA$^{\alpha}$, 
T.~RO{\.Z}EK$^{\beta,\zeta}$, R.~SANTO$^{\gamma}$, T.~SEFZICK$^{\beta}$, M.~SIEMASZKO$^{\zeta}$, J.~SMYRSKI$^{\alpha}$,
A.~T{\"A}SCHNER$^{\gamma}$, P.~WINTER$^{\beta}$, M.~WOLKE$^{\beta,\eta}$, P.~W{\"U}STNER$^{\theta}$, W.~ZIPPER$^{\zeta}$ \\
(COSY-11 Collaboration)}

\vspace{0.2cm}
\address{$^{\alpha}$Institute of Physics, Jagellonian University,
Cracow, PL-30-059, Poland\\
$^{\beta}$IKP, Forschungszentrum J{\"u}lich,  
J{\"u}lich, D-52425, Germany\\
$^{\gamma}$IKP, Westf{\"a}lische Wilhelms-Universit{\"a}t, 
M{\"u}nster, D-48149, Germany\\
$^{\delta}$Institute of Nuclear Physics, 
Cracow, Poland\\
$^{\epsilon}$ Accelerator Division, Gesellschaft f{\"u}r Schwerionenforschung,  
Darmstadt, Germany\\
$^{\zeta}$Institute of Physics, University of Silesia, 
Katowice, Poland\\
$^{\eta}$Svedberg Laboratory, 
Uppsala, Sweden\\
$^{\theta}$ZEL, Forschungszentrum J{\"u}lich,
J{\"u}lich, Germany\\ 
}



\maketitle


\begin{abstract}
A brief experimental overview  
on the close-to-threshold $\eta$ meson production in proton-proton interactions
 is presented and the available observables in measurements with unpolarized and
polarized beam and target are discussed.

\end{abstract}

\section{Measured observables}
The still not well established production mechanism of the $\eta$ meson as well as its
interaction with protons can be investigated via measurements of the following 
observables available in proton-proton scattering: 

\subsection{Total cross section}

Measurements of the total cross section for the $pp\to pp\eta$ reaction have been performed 
at various excess energies in many laboratories worldwide~\cite{eta_exp,calen99,abdel,moskal04,disto}.
The existing data, gathered
utilizing different experimental techniques are in
excellent agreement with each other~\footnote{Due to limited space
we are not presenting here the figure of the excitation function 
for the $pp\to pp\eta$ reaction. For an overview picture see for instance
Fig.~8a in~\cite{moskal02}}. Theoretical considerations of the excitation function 
led in most cases to the conclusion 
that close-to-threshold production of the $\eta$ mesons in proton-proton collisions
proceeds predominantly via the excitation and deexcitation of the negative parity
S$_{11}$(1535) resonance~\footnote{For an overview of   
theoretical models see section 7.3 of~\cite{hanhart} and references therein.}. 
Despite the very 
good agreement between different theoretical models as far as the leading 
$\eta$ creation mechanism is concerned, there are a lot of discrepancies 
when trying to explain which out of the considered exchanged mesons 
plays the dominant role in the excitation of the S$_{11}$. As it was 
shown by Nakayama et al.~\cite{nakayama02}, both pseudoscalar and vector meson exchanges
result in equally well descriptions of the excitation function 
for the $pp\to pp\eta$ reaction close to threshold.

\subsection{Angular distributions}

Measurements of the close-to-threshold angular distributions of 
the $\eta$ meson emission in the centre-of-mass system 
at excess energies of Q=15~\cite{abdel}, 15.5~\cite{moskal04}, 37~\cite{calen99}, and 41~MeV~\cite{abdel}
showed rather flat distributions. It is concluded from these data that in the  
lower excess energy range the $\eta$ meson is mainly produced in the $s$-wave, while 
at higher excess energies the influence of the $p$-wave starts to play an 
important role~\cite{nakayama03}. The contribution from the $d$-wave seems to be negligible. 
Also the proton's angular distributions of the centre-of-mass emission angle~\cite{abdel,moskal04}
show an isotropic behaviour. At Q=15~MeV the data indicate dominance of the $^{3}P_{0}\to ^{1}\!\!S_{0}s$
transition distorted slightly by the contribution from the $^{1}S_{0}\to ^{3}\!\!P_{0}s$.  
At Q=40~MeV there is visible an additional presence of the $^{1}D_{2}\to ^{3}\!\!P_{2}s$ transition~\cite{nakayama03}. 
Contributions from the other
partial waves were found to be  
of minor importance.
In contradiction to these results, the emission plane of the 
$pp\to pp\eta$ reaction was found to be anisotropic~\cite{moskal04}~\footnote{Which is in 
line with the result by~\cite{rodeburg}}. 
This effect has not been explained theoretically yet.

\subsection{Invariant mass distributions}

The high statistics production
of $\eta$ mesons in proton-proton collisions at $Q=15.5$~MeV
allowed to perform a Dalitz plot analysis of the 3-body final state~\cite{moskal04}.
Surprisingly, beside the clear enhancement seen in the lower range of the invariant mass distribution
of the $pp$ subsystem, originating in the strong proton-proton final state interaction, 
the presence of a wide bump in the upper range of this plot
has been observed.  
Different approaches have been applied in order to explain the origin of 
this wide bump. Nakayama et. al.~\cite{nakayama03} 
described the distribution by introducing $P$-waves in the 
$pp$ subsystem. This approach, although in agreement with the shapes
of both $pp$ and $p\eta$ invariant masses, fails when explaining the 
shape of the excitation function for the $pp\to pp\eta$ reaction in 
the range below Q=40~MeV. It is also worth to mention that an indication of such a
bump was also seen in the $pp$ invariant mass distribution at Q=4.5~MeV~\cite{moskal04}, where
one expects the presence of $S$-waves only. A three-body
calculation performed by Fix and Arenh{\"o}vel~\cite{fix} could also describe  
the bump in the $pp$ invariant mass at Q=15.5~MeV well, however it fails in explaning 
the origin of the peak in the low range of $pp$ invariant mass. It also doesn't reproduce the shape 
of the $pp$ invariant mass at Q=41~MeV. The  
$p\eta$ invariant mass distributions were not described in this approach.
Solution based on the parametrization of the reaction amplitude 
proposed by Deloff~\cite{deloff} resulted in a good description of the  
$pp$ and $p\eta$ invariant masses at Q=15.5~MeV. However, 
it fails at Q=40~MeV.  
Another 
model proposed by the same author, based on 
a three-particle pair-wise approach via the hyperspherical 
harmonics~\cite{deloff2}, led to a good description of $pp$ and $p\eta$ invariant mass 
distributions at Q=15.5~MeV. Solutions at Q=40~MeV as well as the
excitation function were not discussed in frame of this model.

\subsection{Analysing power}

Polarisation observables should provide a more detailed information 
on the dynamics of the $\eta$ meson creation in 
hadronic collisions, 
due to their sensitivity to the 
interference terms between $Ps$ and $Pp$ waves. The first attempt 
to measure the proton analysing power for the $\vec{p}p\to pp\eta$ has been undertaken by the COSY-11 group, 
with a set of data at Q=40~MeV~\cite{winter}. The data, within their rather large error bars, 
are consistent with zero, which indicates the absence of higher-than-$s$ partial 
waves in $\eta$ production at Q=40~MeV. The $Ps$ and $Pp$ interference term as well as
the sum of the $(Pp)^{2}$ and $SsSd$ interference terms were extracted in the analysis and
are equal to $(0.003\pm 0.004)\mu$b and $(-0.005\pm 0.005)\mu$b, respectively.
In particular, this may indicate that there is no interference between the $Ps$ and $Pp$ waves.   
Recently, the DISTO collaboration has obtained the set of A$_y$ data at 
excess energies far from threshold, which surprisingly also turned out to be 
consistent with zero within the error bars~\cite{disto}. 

\section{Prospectives}

Beside the upcoming data from the COSY-11 collaboration for the proton analysing power 
in the $\vec{p}p\to pp\eta$ reaction at Q=10 and 37 MeV~\footnote{Status of the 
analysis will be reported in~\cite{czyzyk}.} it was proposed 
to measure the spin correlation function~\footnote{Theoretical predictions for the 
C$_{xx}$ values are given for instance in~\cite{nakayama03,rekalo}.}  
for the $\vec{p}\vec{p}\to pp\eta$ reaction at COSY~\cite{czyzyk2}.
The latter would be possible at an external target station at COSY
with a frozen spin target. For a high efficiencies in the $pp\eta$ 
event separation the WASA detector~\cite{zabierowski}, which will be moved to COSY
in the near future, would be at best suited. The spin correlation function 
is an model-independent direct measure of the contribution of
the $^{3}P_{0}\to ^{1}\!\!S_{0}s$ transition in the creation of the $\eta$ meson~\cite{nakayama03,hanhart}.
Performing this experiment in the range of the excess energies, where the 
only contributions to the production amplitude are those from the
$^{3}P_{0}\to ^{1}\!\!S_{0}s$ and $^{1}S_{0}\to ^{3}\!\!P_{0}s$ transitions, one will be 
able to exctract contributions from the both transitions in this excess energy range.

\section*{Acknowledgments}
The work has been supported by the European Community - Access to
Research Infrastructure action of the
Improving Human Potential Programme,
by the DAAD Exchange Programme (PPP-Polen),
by the Polish State Committe for Scientific Research
(grants No. 2P03B07123 and PB1060/P03/2004/26),
and by the Research Centre J{\"u}lich.


\end{document}